\documentstyle[12pt]{article}
\newcommand\be{\begin{equation}}
\newcommand\ee{\end{equation}}
\newcommand\bea{\begin{eqnarray}}
\newcommand\eea{\end{eqnarray}}

\newcommand{\fatalpha}{{\bf \alpha \kern -0.44em \alpha}}
\newcommand{\fatsigma}{{\bf \sigma \kern -0.54em \sigma}}
\newcommand{\tpchi}{{\bf \chi \kern -0.35em \chi}}
\newcommand{\llambda}{{\bf \lambda \kern -0.45em \lambda}}

\bibliography{plain}
\title{\bf D-concurrence bounds for pair coherent states}\vspace{20mm}
\author{ S. Salimi
  \thanks{Corresponding author:  E-mail addresses:
  shsalimi@uok.ac.ir},  A. Mohammadzade
  \thanks{E-mail addresses: Amir.Mohammadzade@uok.ac.ir}
 \\ {\small Department of Physics,
University of Kurdistan, P.O.Box 66177-15175 , Sanandaj, Iran.}}
\pagebreak

\vspace{20mm}
\begin{document}
\maketitle \vspace{15mm}
\begin{abstract}
The pair coherent state is a state of a two-mode radiation field
which is known as a state with non-Gaussian wave
function. In this paper, the upper and lower bounds for D-concurrence
(a new entanglement measure) have been studied over this state and calculated.
{\bf PACs Index: } \vspace{70mm}
\end{abstract}

\section{Introduction}
Quantum information processing has been the focus of recent
quantum scientific research and has attracted a lot of attention.
Quantum entanglement is one of the key resource for quantum
information processing and manipulating of entangled states are
essential for quantum information applications. Such these
applications are quantum teleportation \cite {1,2}, quantum
cryptography \cite {3,4}, quantum dense coding \cite {5,6,7}, and
quantum computation \cite {8,9,10}. The fundamental question in
quantum entanglement theory is which states are entangled and
which ones are not? Only in some cases we can find the simple
answer to this question. The case of pure bipartite states is the
simplest ones. Any bipartite pure state $|\Psi_{AB}\rangle \in
H_{AB} = H_{A} \otimes H_{B}$ is called separable (entangled) iff
it can be (can not be) written as a product of two vectors
corresponding to Hilbert spaces of subsystems: $|\Psi_{AB}\rangle
= |\phi_{A}\rangle|\psi_{B}\rangle$.
\\Bennett et \textit{al} \cite {11} has defined a measure of entanglement for each
pure state of a bipartite system $|\rho_{AB}\rangle$
as below:

\begin{equation}E(|\rho_{AB}\rangle) = -Tr(\rho_{A}\log_{2}\rho_{A})
= -Tr(\rho_{B}\log_{2}\rho_{B}), \end{equation} which is called as
entropy of entanglement. $\rho_{A} =
Tr_{B}|\rho_{AB}\rangle\langle\rho_{AB}|$ is the partial trace of
$\rho$ over subsystem B, and $\rho_{B}$ has a similar meaning.
Some measures such as concurrence \cite {12,13,14,15}, negativity
\cite {16,17,18}, and tangle \cite {19,20,21} can be used for
quantifying entanglement. Experimental quantification of
entanglement has attracted more attention recently \cite
{22,23,24,25}. D-concurrence is a measures for quantifying the
amount of entanglement which  have proposed by Ma and Zhang \cite
{26}. This measure has advantages in comparison with other
measures, specially concurrence, such as simplicity of form and
accuracy of results.
\\Another important concept which widely used and very useful for studying of different problems
in quantum information theory is coherent states or quasiclassical
states which first introduced by Schr\"{o}dinger in 1926 \cite {27}
and then it extended by Glauber \cite {28} and Perelomov \cite {29}.
Coherent states are applied for study of entangled nonorthogonal
states, also they have vital importance in quantum optic \cite
{30,31} and mathematical physics \cite {29}. Bosonic entangled
coherent state \cite {32}, SU(1,1), and SU(2) coherent states \cite
{33} are typical examples of entangled coherent states. Recently,
much attention has been paid to continuous variable quantum
information processing in which continuous variable type entangled
pure states play a important role \cite {49,50,51,52}. For example,
two-state entangled coherent states are used realize effective
quantum computation \cite {53} and quantum teleportation \cite {54}.
Two-mode squeezed vacuum states have been applied to quantum dense
coding \cite {55}. Therefore, it is an attractive subject to apply
and study continuous variable type entangled pure states. One of
these states is pair coherent state where preliminary concept of
this state was presented by Agarwal \cite {34,35}. Agarwal suggested
that the optical pair coherent state can be generated via the
competition of four-wave mixing and two-photon absorption in a
nonlinear medium. Another scheme has been suggested for generating
vibrational pair coherent states via the motion of a trapped ion in
a two-dimensional trap \cite {36}. Since calculation of entanglement
measures for high dimension state is difficult, it is a urgent task
to find bound for entanglement measures \cite {37,38,39,40,41,42}.
The basic aim of this paper is calculation of upper and lower bounds
of D-concurrence over a family of non-Gaussian states, namely, the
pair coherent state.
\\ The organization of this paper is as follows: In sec. 2, we have
review the pair coherent state, investigated their properties
briefly and at the end, we indicate to Peres-Horodecki criterion.
In sec. 3, we have calculated upper and lower bound of the
D-concurrence over pair coherent state. The conclusion is given
in sec. 4.

\section{Pair coherent state: A state with non-Gaussian wave function}
The pair coherent states (PCS) are regarded as an important type
of correlated two-mode states, which possess prominent
nonclassical properties \cite {43} such as sub-Poissonian
statistics, strong intermode correlation in the number
fluctuations, squeezing of quadrature variances, and violations
of Cauchy-Schwarz inequalities and they have been extensively
studied for violation of Bell inequalities \cite {44,45}. Such
states denoted by $|\zeta,q\rangle$ where are states of a two-mode
radiation field  \cite {34,35} with the following properties
\begin{equation}\matrix{ ab|\zeta,q\rangle=\zeta|\zeta,q\rangle \cr
(a^{\dag}a-b^{\dag}b)|\zeta,q\rangle=q|\zeta,q\rangle,}\end{equation}
where a and b are the annihilation operators associated with two
modes, $\zeta$ is a complex number, and q is the degeneracy
parameter. Pair coherent states can be explicitly expanded as a
superposition of the two-mode Fock states, i.e.,
\begin{equation}|\zeta,q\rangle = N_{q}\sum^{\infty}_{n=0}\frac{\zeta^{n}}{(n+q)!}|n+q,n\rangle,\end{equation}
where the normalization constant $N_{q}$ is given by ($I_{q}$ is
the modified Bessel function of the first kind of order q)
\begin{equation}N_{q} = [\mid \zeta \mid^{-q} I_{q}(2\mid \zeta \mid)]^{\frac{-1}{2}}.\end{equation}
The pair coherent state for q = 0 (corresponding to equal photon
number in both the modes) is given by \cite {56}
\begin{equation}|\zeta,0\rangle = N_{0}\sum^{\infty}_{n=0}\frac{\zeta^{n}}{n!}|n,n\rangle,\end{equation}
subsequently $N_{0} = \frac{1}{\sqrt{I_{0}(2\mid\zeta\mid)}}$ and
$I_{0}(2\mid\zeta\mid)$ is the modified Bessel function of order
zero. The coordinate space wave function is given by

\begin{equation}\matrix{\langle x_{a},x_{b}|\zeta,0\rangle = \cr
N_{0}\sum^{\infty}_{n=0}\frac{\zeta^{n}}{n!}\langle
x_{a}|n\rangle\langle x_{b}|n\rangle =
N_{0}\sum^{\infty}_{n=0}\frac{\zeta^{n}}{n!}\frac{1}{\sqrt{\pi}}\frac{H_{n}(x_{a})
H_{n}(x_{b})}{2^{n}n!}\exp\left[-\frac{x_{a}^{2}+x_{b}^{2}}{2}\right],}\end{equation}
where $\langle x_{a}|n\rangle$ is a harmonic oscillator wave
function given in terms of the Hermite polynomial as
\begin{equation}\langle x_{a}|n\rangle =
\frac{H_{n}(x_{a})\exp(\frac{-x^{2}_{a}}{2})}{(2^{n}n!\sqrt{\pi})^{\frac{1}{2}}}.\end{equation}
It is clear from the Eq. (2-6) that the wave function of the pair
coherent state is non-Gaussian. Now, we investigate the
inseparability of the pair coherent states in light of
Peres-Horodecki criterion, however the state of Eq. (2-5) has an
obvious form of Schmidt decomposition. This reflects the fact
that this state is an entangled state.

\textbf{Perese-Horodecki criterion}
\\Nonseparability for these states has been established using Peres-Horodecki criterion \cite {16,17,46}. The
Peres-Horodecki inseparability criterion is known to be necessary
and sufficient for the ($2 \times 2$) and ($2 \times 3$)
dimensional states, but to be only sufficient for any higher
dimensional states. This criterion states that if the partial
transpose of a bipartite density matrix has at least one negative
eigenvalue, then the state becomes inseparable. The density
matrix $\rho$ corresponding to the state $|\zeta,0\rangle$ (which
is a infinite dimensional state) can be written as

\begin{equation}\rho = \left(\sum^{\infty}_{n=0}C_{nn}|n,n\rangle\right)
\left(\sum_{m=0}^{\infty}C_{mm}^{*}\langle
m,m|\right),\end{equation} where
$C_{mm}=N_{0}\frac{\zeta^{m}}{m!}$. Partial transpose of Eq. (2-5)
was shown to have negative eigenvalues and therefore the
nonseparability

\begin{equation}\lambda_{nn} =
\frac{1}{I_{0}(2\mid\zeta\mid)}\frac{\mid \zeta
\mid^{2n}}{(n!)^{2}},\forall n
\end{equation}

\begin{equation}\lambda_{nm}^{\pm} =
\pm \frac{1}{I_{0}(2\mid\zeta\mid)}\frac{\mid \zeta
\mid^{n+m}}{(n!m!)},\forall n\neq m.\end{equation} One can in fact
construct negativity $N(\rho)$ by finding absolute sum of negative
eigenvalues in lieu of a computable measure of entanglement as

\begin{equation}N(\rho) =
 \mid \sum^{\infty}_{n=0}\sum_{m=0}^{\infty} \lambda^{-}_{nm}\mid
 = \mid 1 - \frac{e^{2\mid\zeta\mid}}{I_{0}(2\mid\zeta\mid)} \mid, \forall
 n \neq m.
\end{equation}
In the limit $\mid\zeta\mid \rightarrow 0$, $N(\rho)\rightarrow 0
$ which is indicating that there is no entanglement, because in
this limit only $|0,0\rangle$ state will survive in Eq. (2-5) and
all such states like $|n,n\rangle$ are separable.

\section{D-concurrence}
Recently, the novel measure named D-concurrence, has been
proposed by Ma and Zhang, which in comparison with concurrence,
has a advantages such as simplicity of its structure and accuracy
of results \cite {26}. For a mixed state, D-concurrence is defined
by the convex roof, that is, defined as the average D-concurrence
of the pure states of the decomposition, minimized over all
decompositions of $\rho =
\sum_{i}|\psi_{i}\rangle\langle\psi_{i}|$
\begin{equation}D(\rho) = \inf\sum_{i}p_{i}D(\psi_{i}), \end{equation}
where $p_{i}$ are real numbers which satisfy the following
condition
\begin{equation}\sum_{i}p_{i} = 1, \end{equation}
and $\psi_{i}$ are pure states. Upper and lower bounds of
D-concurrence defined by
\begin{equation}[\det (I - \rho_{A}) - \det (I - \rho)] \leq D^{2}(\rho) \leq [\det (I - \rho_{A})],\end{equation}
where $\rho_{A} = Tr_{B}\rho$ is the partial trace of $\rho$ over
subsystem B. It is too difficult to calculate the measure of
entanglement for mixed states in high dimensions, therefore it is
too vital to find the bounds for measures such as D- concurrence
over states like pair coherent states. According to density matrix
$\rho$ in Eq. (2-8), we calculate upper and lower bounds of
D-concurrence over pair coherent states in infinite dimensions.

\textbf{Calculation of upper bound}
\\ First, we calculate upper bound i.e, $\det(I-\rho_{A})$.
According to Eq. (2-8), $\rho_{A}$ is as following
\begin{equation} \rho_{A} = \sum_{n}^{\infty} \mid C_{nn} \mid^{2} |n \rangle\langle n|.\end{equation}
Therefore $I - \rho_{A}$ equals to
\begin{equation}I - \rho_{A} = \sum_{n}^{\infty}(1 - \mid C_{nn} \mid^{2}) |n \rangle\langle n|.\end{equation}
Eq. (3-16) is a diagonal matrix, so that its determinant is
\begin{equation}\det(I - \rho_{A}) = \prod_{n}(1-\mid C_{nn} \mid^{2}),\end{equation}
this equation is upper bound of D-concurrence.

\textbf{Calculation of lower bound}
\\ According to Eq. (3-14), to calculate the lower bound, $[\det (I - \rho_{A}) - \det (I - \rho)]$
should be calculate ($\det (I - \rho_{A})$ has been calculated
above). Elements of the matrix $A=I - \rho$ are as following
\[
A_{i j} = \left\{
\begin{array}{ll}
1-\mid C_{ii} \mid^{2} & \mbox{if $i=j$}\\
-C_{ii}C_{jj}^{*} & \mbox{otherwise.}
\end{array}
\right.
\]
Because $I - \rho$ is a matrix with infinite dimension, therefore it
is too difficult to calculate its determinant. Here, to calculate
$\det(I-\rho)$, first we should calculate determinant of matrixs
$I-\rho$ with low dimensions such as $2 \times 2$, $3 \times 3$, $4
\times 4...$ and then generalize it to high dimensions that process
is as following
\begin{equation} \matrix{\det(I-\rho)_{(2 \times 2)} = 1 - \mid C_{00} \mid^{2} - \mid C_{11}\mid^{2} =
 1 - \sum_{n=0}^{1}\mid C_{nn} \mid^{2} \cr
 \det(I-\rho)_{(3 \times 3)} = 1 - \mid C_{00} \mid^{2} - \mid C_{11}\mid^{2} - \mid C_{22}\mid^{2} =
 1 - \sum_{n=0}^{2}\mid C_{nn} \mid^{2} \cr
 \det(I-\rho)_{(4 \times 4)} = 1 - \mid C_{00} \mid^{2} - \mid C_{11}\mid^{2} - \mid C_{22}\mid^{2} - \mid C_{33}\mid^{2} =
 1 - \sum_{n=0}^{3}\mid C_{nn} \mid^{2} \cr
 .
 \cr
 .
 \cr
 .
 \cr
 \det(I-\rho)_{(N \times N)} = 1 - \mid C_{00} \mid^{2} - \mid C_{11}\mid^{2} - \mid C_{22}\mid^{2}
 - ...- \mid C_{N-1,N-1}\mid^{2} =1 -  \sum_{n=0}^{N-1} \mid C_{nn} \mid^{2}.}\end{equation}
 Therefore determinant of $(I-\rho)$ is
\begin{equation}\det (I-\rho) =  1 - \sum_{n=0}^{N-1}\mid C_{nn} \mid^{2},\end{equation}
where $N =2, 3, ...\infty$.
 So, considering Eq. (3-14), the lower
bound of D-concurrence is as following
\begin{equation}\prod_{n}(1-\mid C_{nn} \mid^{2}) -1+ \sum_{n=0}^{N-1}\mid
C_{nn} \mid^{2}.\end{equation} At the end, lower and upper bounds
of D-concurrence, for the pair coherent states, are
\begin{equation} (\prod_{n}(1-\mid C_{nn} \mid^{2}) -1+
\sum_{n=0}^{N-1}\mid C_{nn} \mid^{2}) \leq D^{2}(\rho) \leq
(\prod_{n}(1-\mid C_{nn} \mid^{2}))
\end{equation}
where clearly
\begin{equation}\sum_{n=0}^{N-1}\mid C_{nn} \mid^{2} < 1.\end{equation}
\section{Conclusion}
In this paper we focus on the family of non-Gaussian states that
are known as continuous variable or infinite dimensional system,
and we have studied measure of D- concurrence on these states.
Since the D-concurrence for high dimension mix state is difficult
to calculate, it is a necessary to find bound for D-concurrence,
hence we have computed upper and lower bounds of D-concurrence
over the pair coherent states.

\newpage
{\bf }
\end{document}